\documentstyle [12pt]{article}  
\begin{document}

\begin{center}
              
{\bf Measure Fields, The Cosmological Constant and Scale Invariance}                   

\end{center}                                                                  

\bigskip                                                                      

\begin{center}                                                                      
                        
E. I. Guendelman\footnote{Electronic address: guendel@bgumail.bgu.ac.il}      
                                                                              
{\it Physics Department, Ben Gurion University, Beer Sheva, Israel}           
\end{center}                                                                  
                                                                              
\bigskip                                                                      

\begin{abstract}
The consequences of considering the measure of 
integration in the action to be defined by degrees of freedom independent
of the metric are studied. Models without the cosmological constant problem,
new ways of spontaneously breaking scale symmetry which have an interesting 
cosmology and theories of extended objects (string, branes) without a 
fundamental scale appear possible.
\end{abstract}

When formulating generally covariant theories, the form    
for the action $S = \int d^{D}x \sqrt{-g}L$ is considered. Here $L$ is a scalar
and $g = det(g_{\mu \nu})$, $g_{\mu \nu}$ being the metric defined on 
space-time manifold of interest. In this case the metric determines the 
measure of integration to be $d^{D}x \sqrt{-g}$. 
The fact is however that one can still maintain general coordinate     
invariance and replace $\sqrt{-g}$ by another density $\Phi$, which    
depend on degrees of freedom independent of the metric, obtaining      
then the modified measure $d^{D}x \Phi $. If $\Phi$ is also a total    
divergence, then the action $S = \int d^{D}x \Phi L$ is invariant      
up to the integral of a total divergence under the shift               
$L \rightarrow L + constant$.

A particular realization for $\Phi$ is obtained by considering $D$     
scalar fields $\varphi_{j}$ and defining                               
$\Phi=\varepsilon^{\mu_{1}...\mu_{D}}\varepsilon_{j_{1}...j_{D}}       
\partial_{\mu_{1}}\varphi_{j_{1}}....\partial_{\mu_{D}}\varphi_{j_{D}}$.

Realistic gravitational theories without the cosmological constant problem 
can be constructed using actions of the form            
$S =\int d^{D}x \Phi L_{1} + \int d^{D}x \sqrt{-g} L_{2} $, where      
$L_{1}$ and $L_{2}$ are both independent of the measure fields \cite{cosm}         
$\varphi_{j}$. 
There is a good reason not to consider            
mixing of  $\Phi$ and                             
$\sqrt{-g}$ , like                                                            
for example using                                 
$\frac{\Phi^{2}}{\sqrt{-g}}$. This is because $S$ is invariant 
(up to the integral of a  divergence) under the infinite dimensional symmetry                           
$\varphi_{a} \rightarrow \varphi_{a}  +  f_{a} (L_{1})$                       
where $f_{a} (L_{1})$ is an arbitrary function of $L_{1}$ if $L_{1}$ and     
$L_{2}$ are $\varphi_{a}$    independent. Such symmetry is
absent if mixed terms are present. Models with two measures could be
related to brane scenarios (one measure associated to the brane, the 
other to the bulk).                                            
                                                                              
        We will study now the dynamics of a scalar field $\phi$ interacting   
with gravity as given by the choice of lagrangians (for $D=4$)  \cite{dil}                              
                                         
$L_{1} = \frac{-1}{\kappa} R(\Gamma, g) + \frac{1}{2} g^{\mu\nu}               
\partial_{\mu} \phi \partial_{\nu} \phi - V(\phi),  L_{2} = U(\phi)$                                                                           
where                                      
$R(\Gamma,g) =  g^{\mu\nu}  R_{\mu\nu} (\Gamma) , R_{\mu\nu}                   
(\Gamma) = R^{\lambda}_{\mu\nu\lambda}, R^{\lambda}_{\mu\nu\sigma} (\Gamma) = 
\Gamma^{\lambda}_{\mu\nu,\sigma} - \Gamma^{\lambda}_{\mu\sigma,\nu} +                          
\Gamma^{\lambda}_{\alpha\sigma}  \Gamma^{\alpha}_{\mu\nu} -                   
\Gamma^{\lambda}_{\alpha\nu} \Gamma^{\alpha}_{\mu\sigma}$.                     
In the variational principle $\Gamma^{\lambda}_{\mu\nu},              
g_{\mu\nu}$, the measure fields scalars                                       
$\varphi_{a}$ and the  scalar field $\phi$ are all to be treated              
as independent variables.

        If we perform the global scale transformation ($\theta$ =             
constant)                    
 $g_{\mu\nu}  \rightarrow   e^{\theta}  g_{\mu\nu}$                                                                
then $S$ is invariant provided  $V(\phi)$     
and $U(\phi)$ are of the                                                      
form                                                                          
$V(\phi) = f_{1}  e^{\alpha\phi},  U(\phi) =  f_{2}                            
e^{2\alpha\phi}$                                                                
and $\varphi_{a}$ is transformed according to                                 
$\varphi_{a}   \rightarrow   \lambda_{a} \varphi_{a}$                         
(no sum on a) which means                                                     
$\Phi \rightarrow \biggl(\prod_{a} {\lambda}_{a}\biggr) \Phi \equiv \lambda   
\Phi $                                                                        
such that                                                                     
$\lambda = e^{\theta}$                                                        
and                                                                           
$\phi \rightarrow \phi - \frac{\theta}{\alpha}$.

        Let us consider the equations which are obtained from                 
the variation of the $\varphi_{a}$                                            
fields. We obtain then  $A^{\mu}_{a} \partial_{\mu} L_{1} = 0$               
where  $A^{\mu}_{a} = \varepsilon^{\mu\nu\alpha\beta}                         
\varepsilon_{abcd} \partial_{\nu} \varphi_{b} \partial_{\alpha}               
\varphi_{c} \partial_{\beta} \varphi_{d}$. Since                              
det $(A^{\mu}_{a}) =\frac{4^{-4}}{4!} \Phi^{3}$, 
if $\Phi\neq 0$ we obtain that $\partial_{\mu} L_{1} = 0$,          
 or that                                                                      
$L_{1}  = M$,                                                                 
where $M$ is constant. This constant $M$ appears in a self-consistency            
condition of the equations of motion \cite{dil}                                         
that allows us to solve for $ \chi \equiv \frac{\Phi}{\sqrt{-g}}
 = \frac{2U(\phi)}{M+V(\phi)}$. $M$ produces ssb of scale invariance.

        To get the physical content of the theory, it is convenient to go     
to the Einstein conformal frame where       
$\overline{g}_{\mu\nu} = \chi g_{\mu\nu}$ . 
In terms of $\overline{g}_{\mu\nu}$   the non       
Riemannian contribution (defined   as                                         
$\Sigma^{\lambda}_{\mu\nu} =                                                  
\Gamma^{\lambda}_{\mu\nu} -\{^{\lambda}_{\mu\nu}\}$                           
where $\{^{\lambda}_{\mu\nu}\}$   is the Christoffel symbol),                 
disappears from the equations, which can be written then in the Einstein      
form ($R_{\mu\nu} (\overline{g}_{\alpha\beta})$ =  usual Ricci tensor)        

$R_{\mu\nu} (\overline{g}_{\alpha\beta}) - \frac{1}{2}                         
\overline{g}_{\mu\nu}                                                         
R(\overline{g}_{\alpha\beta}) = \frac{\kappa}{2} T^{eff}_{\mu\nu}(\phi)$,                                                                                                                                                  
                                                              
$T^{eff}_{\mu\nu} (\phi) = \phi_{,\mu} \phi_{,\nu} - \frac{1}{2} \overline     
{g}_{\mu\nu} \phi_{,\alpha} \phi_{,\beta} \overline{g}^{\alpha\beta}          
+ \overline{g}_{\mu\nu} V_{eff} (\phi)$ and                                       
$V_{eff} (\phi) = \frac{1}{4U(\phi)}  (V+M)^{2}$. Notice that for generic 
smooth functions $V$ and $U$, if $V+M = 0$ at some point, then generically
$V_{eff} = V^{'}_{eff} = 0$ at such point, that is, the ground state has zero vacuum 
energy without fine tuning!. No assumption of scale invariance in involved
in this result.

        If $V(\phi) = f_{1} e^{\alpha\phi}$  and 
 $U(\phi) = f_{2}e^{2\alpha\phi}$ as                                                           
required by scale invariance, we obtain 
\begin{equation}                                                              
        V_{eff}  = \frac{1}{4f_{2}}  (f_{1}  +  M e^{-\alpha\phi})^{2}        
\end{equation}

We see that as $\alpha \phi \rightarrow  \infty$,
$V_{eff} \rightarrow \frac{f_{1}^{2}}{4f_{2}} = const$.                
providing an infinite flat region, so we expect a slow  rolling  
inflationary scenario to be viable. Also a minimum is achieved at zero         
cosmological constant for the case $\frac{f_{1}}{M} < 0 $ at the point         
$\phi_{min}  =  \frac{-1}{\alpha} ln \mid\frac{f_1}{M}\mid $. Finally,        
the second derivative of the potential  $V_{eff}$  at the minimum is          
$V^{\prime\prime}_{eff} = \frac{\alpha^2}{2f_2} \mid{f_1}\mid^{2} > 0 $        
if $f_{2} > 0$. That is, a realistic scalar field potential, with           
massive excitations when considering the true vacuum state, is achieved in    
a way consistent with scale invariance.                     
                                                                              
            Furthermore, one can consider this model as suitable for the      
present day universe rather than for the early universe, after we suitably    
reinterpret the meaning of the scalar field  $\phi$. This can provide a long  
lived almost constant vacuum energy for a                                     
long period of time, which can be small if $f_{1}^{2}/4f_{2}$ is              
small. Such small energy                                                      
density will eventually disappear when the universe achieves its true         
vacuum state. Notice that a small value of $\frac{f_{1}^{2}}{f_{2}}$   can be       
achieved if we let $f_{2} >> f_{1}$. In this case                             
$\frac{f_{1}^{2}}{f_{2}} << f_{1}$, i.e. a very small scale for the           
energy                                                                       
density of the universe is obtained by the existence of a very high scale     
(that of $f_{2}$) the same way as a small neutrino mass is obtained in the    
see-saw mechanism from the existence also of a large mass scale. In     
what follows, we will take $f_{2} >> f_{1}$.                                                                                                     
                                                                              
We can also include  a fermion $\psi$, where the kinetic term     
of the fermion is chosen to be part of $L_1$, $S_{fk} = \int L_{fk} \Phi d^4 x$
and
$L_{fk} = \frac{i}{2} \overline{\psi} [\gamma^a V_a^\mu                        
(\overrightarrow{\partial}_\mu + \frac{1}{2} \omega_\mu^{cd} \sigma_{cd})     
- (\overleftarrow{\partial}_\mu + \frac{1}{2} \omega_\mu^{cd} \sigma_{cd})    
\gamma^a V^\mu_a] \psi $                                                                
there $V^\mu_a$ is the vierbein, $\sigma_{cd}$ =                              
$\frac{1}{2}[\gamma_c,\gamma_d]$. For            
self-consistency, the curvature scalar is taken to be (if we want to deal     
with the spin connection $\omega_\mu^{ab}$ instead of $\Gamma^\lambda_{\mu\nu}$ everywhere)       
$R = V^{a\mu}V^{b\nu}R_{\mu\nu ab}(\omega),                                    
R_{\mu\nu ab}(\omega)=\partial_{\mu}\omega_{\nu ab}                           
-\partial_{\nu}\omega_{\mu ab}+(\omega_{\mu a}^{c}\omega_{\nu cb}             
-\omega_{\nu a}^{c}\omega_{\mu cb})$.                                                               
                                                                              
Global scale invariance is obtained                                           
provided $\psi$ also transforms, as in                                        
$\psi \rightarrow \lambda ^{-\frac{1}{4}} \psi$. Mass terms consistent with    
scale invariance exist: $m_1\int \overline{\psi} \psi e^{\alpha\phi/2} \Phi d^4x 
+ m_2\int \overline{\psi} \psi e^{3\alpha\phi/2} \sqrt{-g} d^4x $. 

If we consider the situation where                                            
$m_1 e^{\alpha\phi/2} \overline{\psi}\psi$                                    
or $m_2 e^{3\alpha\phi/2} \overline{\psi}\psi$ are                            
much bigger than $V(\phi) + M$, i.e. a high density approximation, we         
obtain that the consistency condition is                                                                        
$\chi = -\frac{3m_2}{m_1} e^{\alpha\phi}$. Using this, we obtain,     
after going to the Conformal Einstein Frame (CEF), which involves the 
transformations to the scale invariant fields                                                                      
$\overline V_\mu^a$ = $\chi^\frac{1}{2} V_\mu^a$ and $\psi ^\prime$ =        
$\chi ^{-\frac{1}{4}} \psi$ and they lead to a mass term,                     
\begin{equation}                                                               
-2m_2 ( \frac {|m_1|}{3|m_2|})^{3/2}  \int\sqrt{-\overline {g}}      
\overline{\psi} ^{\prime} \psi ^{\prime} d^4x                                 
\end{equation}                                                                
                                                                              
The $\phi$ dependence of the mass term has disappeared, i.e. masses are       
constants. Low density of matter can also give results  
which are similar to those obtained in the high density approximation, in     
that the coupling of the $ \phi $ field disappears and that the mass term     
becomes of a conventional form in the CEF.               
 In this is the case, we study the limit                      
$\alpha \phi \rightarrow \infty$ . Then $U(\phi) \rightarrow \infty$ and             
$V(\phi) \rightarrow \infty$. In this case, taking                            
$m_1 e^{\alpha\phi/2} \overline{\psi}\psi$                                    
and $m_2 e^{3\alpha\phi/2} \overline{\psi}\psi$                               
much smaller than $V(\phi)$ or $U(\phi)$ respectively, we get                                                   
$\chi = \frac{2f_2}{f_1} e^{\alpha\phi}$. If this is used,        
we obtain the mass term $ m \int\sqrt{-\overline {g}}                                  
\overline{\psi} ^{\prime} \psi ^{\prime} d^4x$, where                         
                                                                              
\begin{equation}                                                              
 m = m_1(\frac {f_1}{2f_2})^{\frac{1}{2}} + 
 m_2(\frac {f_1}{2f_2})^{\frac{3}{2}}
\end{equation}                                                                
                                                                              
The mass term is independent of $\phi$ again. So the interaction of $\phi$
with matter dissapears in the CEF due to scale invariance. This is an explicit
realization of the symmetry which Carroll was looking for avoiding interactions 
of the 'quintessential scalar' and the rest of the world \cite{ca} and which
would apply if we consider the universe being totally in the the region 
$\alpha \phi \rightarrow \infty$. The effects of terms which explicitly 
(rather than spontaneously) break scale invariance have also been studied 
\cite{ka}.

Other scenarios can be explored: taking  $m_1$ and $m_2$ of the same               
order of magnitude, we see that the mass of the Dirac particle is much        
smaller in the region $\alpha \phi \rightarrow \infty$, for which (3) is            
valid, than it is in the region of high density of the Dirac particle         
relative to $V(\phi)+M$, as displayed in eq. (2), if the "see-saw"           
assumption $\frac{f_1}{f_2} < < 1$ is made.                                   
Therefore if space is populated by diluted Dirac particles of this      
type, the mass of these particles will grow substantially if we go to the true
vacuum valid in the absence of matter, i.e. $V+M=0$.                                                
In the region $\alpha \phi \rightarrow \infty$, we minimize the       
matter energy, but maximize the potential energy $V_{eff}$ and at        
$V+M=0$, we minimize $V_{eff}$, and particle masses are big. The true vacuum 
state must be in a balanced intermediate stage. Clearly how much above $V+M=0$    
such true vacuum is located must be correlated to how much particle density   
is there in the Universe. A non zero vacuum energy, which must be of the      
same order of the particle energy density, has to appear and this could       
explain the "accelerated universe" that appears to be implied by the most     
recent observations, together with the "cosmic coincidence", that requires                  
the vacuum energy be comparable to the matter energy.

One can also develop a modified measure theory for the case         
of extended objects \cite{ext} including  super symmetric               
strings and branes \cite{sext}. In the case of strings, we can replace in the  
Polyakov action, the measure        
$\sqrt{-\gamma}d^{2}x$ (where $\gamma_{ab}$ is the metric              
defined on the world sheet, $\gamma = det (\gamma_{ab})$ and           
$a,b$ indices for the world sheet coordinates) by $\Phi d^{2}x$,       
where  $\Phi = \varepsilon^{ab} \varepsilon_{ij}                       
\partial_{a} \varphi_{i}\partial_{b} \varphi_{j}$. Then for the        
bosonic string , we consider the action                             
                                                                       
\begin{equation}                                                       
S = - \int d\tau d\sigma \Phi[                                         
\gamma^{ab} \partial_{a} X^{\mu}\partial_{b} X^{\nu} g_{\mu \nu}       
- \frac {\varepsilon^{ab}}{\sqrt{-\gamma}} F_{ab} ]                    
\nonumber                                                              
\end{equation}

where $ F_{ab} = \partial_{a} A_{b} - \partial_{b} A_{a} $ and 
$A_{a}$ is a gauge field defined in the world sheet of the string.
The term with the gauge fields is irrelevant if the ordinary                  
measure of integration is used, since in that case it would be a divergence,  
but is needed for a consistent dynamics in the modified measure reformulation 
of string theory. This is due to the fact that if we avoid such a             
contribution to the action, the variation of the             
action with respect to $\gamma^{ab}$ leads                                    
to the vanishing of the induced metric on the string.                         
The equation of motion obtained from the variation of the                     
gauge field $A_{a}$ is                                                        
$\varepsilon^{ab}\partial_{a} (\frac{\Phi}{\sqrt{-\gamma}}) = 0$ . From which 
we obtain that $\Phi = c \sqrt{-\gamma}$ where $c$ is a constant which can be 
seen is the string tension. The string tension appears then as an integration 
constant and does not have to be introduced from the beginning. The string    
theory Lagrangian in the modified measure formalism does not have any         
fundamental scale associated with it.                                                                               
Extensions to both the super symmetric case \cite{sext} and to  higher branes         
( see both Refs. \cite{ext} and  \cite{sext}) are possible.

\end{document}